# Magnetic properties and enhanced magnetocaloric effect in EuAl$_3$Si single crystals

Hai Zeng*, Shuo Zou, Zhou Wang, Ziyu Li, Kangjian Luo and Yongkang Luo*

*Wuhan National High Magnetic Field Center and School of Physics,*

*Huazhong University of Science and Technology, Wuhan 430074, China*

**ABSTRACT**

This study presents systematic investigations into the growth and physical properties of EuAl$_3$Si single crystals, encompassing magnetic, transport, and thermodynamic analyses. EuAl$_3$Si undergoes a ferromagnetic transition at $T_C$ = 15 K. A significant reversible magnetocaloric effect was observed around $T_C$. Strikingly, with a small change of magnetic field 2 T, the maximum values of magnetic entropy change (13.4 J/kg K), refrigerant capacity (166 J/kg) and adiabatic temperature change (7.2 K) are found. These parameters respectively are 60%, 148% and 64% larger than those of the parent compound EuAl$_4$, and suggest EuAl$_3$Si as an excellent candidate for magnetic refrigeration applications near the temperature of liquid hydrogen. The possible mechanism for this enhancement is also discussed.

Corresponding authors:

* E-mail addresses: mpzslyk@gmail.com (Yongkang Luo) and zenghai320@sina.com (Hai Zeng)

## I. INTRODUCTION

Solid-state cooling based on the magnetocaloric effect (MCE) is garnering increasing attention due to its high energy efficiency, operational stability, and environmental sustainability.[1-3] When a magnetic material is magnetized or demagnetized, its magnetic entropy changes, which results in release or absorption of heat. By exploiting MCE effect, adiabatic demagnetization has become one of the mainstream methods for achieving milli-Kelvin ultralow temperatures.[4] The magnetic entropy change ($\Delta S_M$) is one of the most important parameters in characterizing the MCE of materials. Many new magnetic materials with significant MCE have been widely researched for this purpose, especially for below the temperature region of liquid nitrogen[5-8]. However, the performance of most of these materials requires improvement for magnetic refrigeration applications, as they either necessitate a large operational field or exhibit significant hysteresis and aging effects associated with first-order magnetic transitions or magneto-structural coupling.[9-11] Thus, new magnetic-refrigeration materials with reversible MCE, low operation field and high performance are highly demanded.

Rare-earth (*RE*)-based alloys have gathered much attention owing to the considerable MCE induced by low magnetic field.[12-19] Among them, $Eu^{2+}$ ($4f^7$, $S = 7/2$), which can be regulated easily by magnetic field, pressure or temperature, are a good platform for studying various exotic phenomena such as large negative magnetoresistance, pressure-induced superconducting state, and the magnetocaloric effect, etc.[17,20-21] Recently, $EuAl_4$ was extensively investigated due to its rich and unique magnetic and electronic behaviors such as three successive anti-ferromagnetic (AFM) transitions within a charge-density-wave (CDW) ordered state,[22-25] possible topological Hall effect[26] and magnetic skyrmions.[27] Moreover, a moderate $-\Delta S_M$ =8.4 J/kg K was reported under a low operation field 2 T around 15.8 K,[17] implying a possible candidate for magnetic-refrigeration materials at hydrogen temperature. An intuitive thinking is that proper substituting Al with Si (electron doping) may modify the RKKY (Ruderman-Kittel-Kasuya-Yosida) exchange,[28] and switches the AFM order into a ferromagnetic (FM) order, which potentially will speed up the demagnetization process and thus enhance the magnetic-refrigeration performance. Although polycrystalline sample growth of $EuAl_{4-x}Si_x$ has been reported decades ago,[29] little has been known about its physical properties, yet.

In this work, we present a detailed study on the magnetic properties, MCEs and magnetotransport of $EuAl_3Si$. A ferromagnetic-like order is observed below 15 K for Eu moments. Most importantly, the MCE around 15 K is found to be reversible without magnetic / thermal hysteresis. The excellent MCE is manifested by the maximum values of magnetic entropy change (13.4 J/kg K), refrigerant capacity (166 J/kg) and adiabatic temperature change (7.2 K) that are achieved under a small operation field of 2 T. These values are 1.6, 2.48 and 1.64 times that of the undoped $EuAl_4$, respectively. These parameters indicate that $EuAl_3Si$ may be an effective material for magnetic refrigeration at liquid hydrogen temperatures.

## II. EXPERIMENTAL METHODS

$EuAl_3Si$ single crystals were grown using Al self-flux method. High-purity starting

materials Eu grains (Energy Chemical, 99.9%), Al shots (Al (Alfa Aesar, 99.9999%) and Si powder (Alfa Aesar, 99.9999%) were weighed in a molar ratio of 1:20:1. The mixture was loaded into an alumina crucible, which was subsequently sealed in an evacuated quartz tube. The tube was heated to 1175 °C in 20 h, held for 24 h, and then slowly cooled down to 750 °C at a rate of 2 °C/h, at which the Al flux was removed by centrifugation.

The composition of the compound was confirmed by energy-dispersive X-ray spectroscopy (EDS). The crystal structure and crystallographic orientations were determined at room temperature by a PANalytical x-ray diffractometer (XRD) with Cu $K_{\alpha 1}$ ($\lambda_{K\alpha 1}$ = 1.5406 Å) radiation and Rigaku XtaLAB mini II single crystal XRD with Mo radiation ($\lambda_{K\alpha}$ = 0.71073 Å), respectively. Rietveld refinements were carried out to obtain its structural information. Electrical resistivity and specific heat data were measured by using a physical property measurement system (PPMS-9, Quantum Design). Magnetizations were collected by a magnetic property measurement system (MPMS, Quantum Design). Measurements of the temperature dependence of magnetization and resistivity were performed at a rate of 3 K/min.

III. RESULTS AND DISCUSSION

EuAl$_3$Si adopts the BaAl$_4$-type centrosymmetric tetragonal structure (*I*4/*mmm*, No.139), as illustrated in Fig.1(a). Eu atoms locate at the 2*a* (0, 0, 0) site, 2/3 of Al atoms (Al1) occupy the 4*d* (0.5, 0, 0.25) site, while the rest of Al atoms (Al2) and Si atoms are randomly distributed at the 4*e* (0, 0, 0.3857) site. Figure 1(b) represents the EDS results for the sample, which gives the elemental proportions Eu : Al : Si = 1 :

3.02(5) : 1.04(2), close to the nominal composition. Figure 1(c) shows the Rietveld refinement of the XRD pattern. No extraneous phases can be identified. The lattice parameters obtained are $a = b = 4.410(1)$ Å, $c = 10.852(4)$ Å, in good agreement with earlier polycrystalline results in the literature.[29] The photograph of single crystals are plate-like, typically about $3\times1.5\times0.8$ mm$^3$ as shown in the inset. In Fig. 1(d), only (0 0 2$l$) reflections are visible, indicating that the $c$-axis is perpendicular to the crystal plane. The full-width at half-maximum (FWHM) of the (002) peak is only 0.042°, indicating the high quality of single crystal.

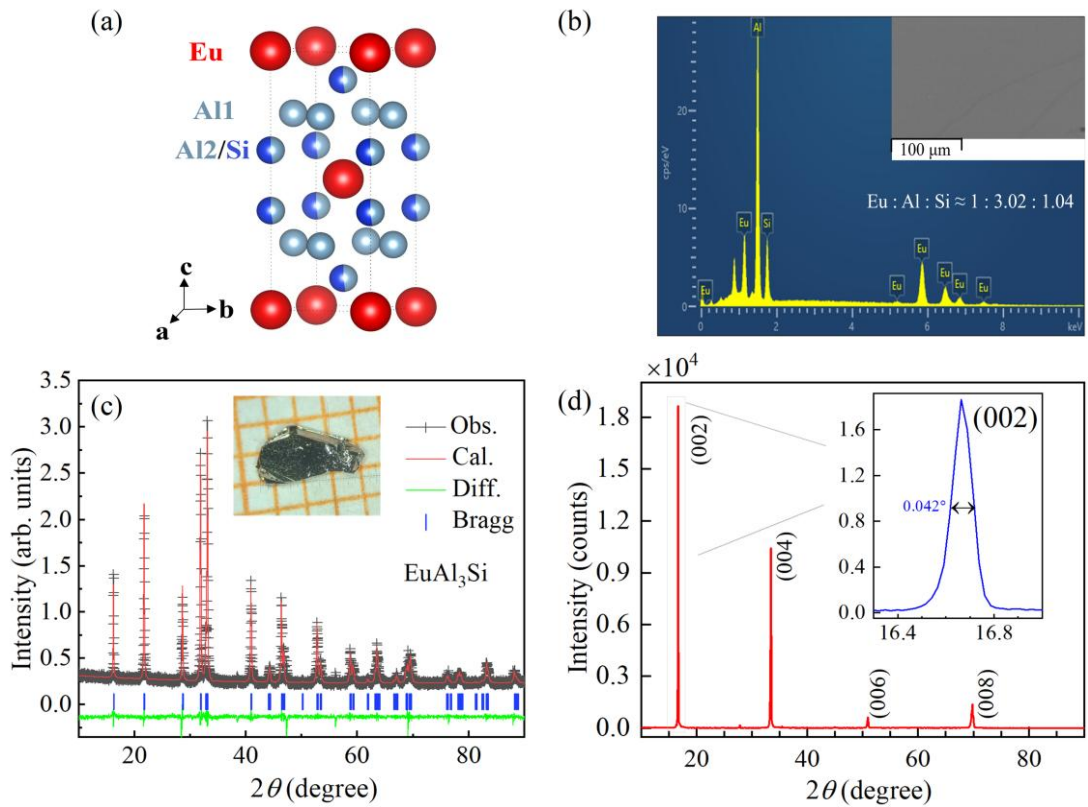

**FIG. 1** (a) Schematic diagram of the crystal structure of EuAl$_3$Si; red, cinereous, and blue balls represent Eu, Al, and Si atoms, respectively. (b) EDS for the crystal. (c) The observed and refined XRD patterns of EuAl$_3$Si powders; inset shows the photograph of a representative single crystal on millimeter-grid paper. (d) The single-crystal XRD pattern; inset gives the zoom-in plot for the (002) peak.

Figure 2(a) shows the temperature dependence of magnetic susceptibility at $\mu_0H =$

0.1 T. At high temperature, both $\chi_{ab}$ and $\chi_c$ well obey the Curie-Weiss law $\chi = C_0 / (T - \theta_{cw})$, where $C_0$ and $\theta_{cw}$ are the Curie constant and paramagnetic Curie temperature, respectively. To determine the effective moments ($\mu_{eff}$) and $\theta_{cw}$, the powder-averaged magnetic susceptibility is calculated, $\chi_{avg} = (\chi_c + 2\chi_{ab})/3$, the inverse of which is shown in the inset to Fig. 2(a) as a function of $T$. The Curie-Weiss fitting of $\chi_{avg}(T)$ leads to $\mu_{eff} = 7.99$ u$_B$/Eu, and $\theta_{cw} = 14.36$ K. The effective moment is close to the theoretical values for Eu$^{2+}$ ($g_J\sqrt{J(J+1)}\mu_B = 7.94\ \mu_B$, $J = S = 7/2$, $g_J = 2$). The positive $\theta_{cw}$ confirms the FM-dominated correlation. Below ~ 15 K, $\chi_{ab}$ increases rapidly before saturating in the $T\rightarrow 0$ limit. On the contrary, $\chi_c$ shows a shallow peak near 15 K below which it tends to level off, and its value is also much smaller than $\chi_{ab}$. Such a behavior indicates that the Eu moments undergo a FM transition near $T_C = 15$ K, and the magnetic moments exhibit a propensity for alignment within the $ab$ plane. Typically, the anisotropy for an ion with $J = S = 7/2$ is notably minimal. This discrepancy could stem from the anisotropic magnetic exchange, and this is partly demonstrated by their different paramagnetic Curie temperatures, $\theta_{CW}^{ab} = 20.7$ K, and $\theta_{CW}^{c} = -8.2$ K. It should be mentioned that the shallow peak observed in $\chi_c(T)$ may also hint that a small component of AFM order may exist along $c$ axis. This can be confirmed by magnetotransport measurements discussed later on. In addition, no magnetic hysteresis is detectable in the whole temperature for both field directions, reaffirming a second-order phase transition.

The magnetization curves $M(H)$ rise rapidly and reach saturation gradually below $T_C$ in both directions at 2 K as shown in Fig. 2(b), consistent with a FM-dominated behavior. The saturated moment is $M_S = 6.75\ \mu_B$ per $Eu^{2+}$ above 0.8 T for **H**//**ab** plane. A larger critical field $H_C \sim 2$ T is needed to saturate for **H**//**c**, while the saturated moment is similar. Note that the FM transition in EuAl$_3$Si is of second order nature, as manifested by the λ-shaped peak in specific heat shown in the inset to Fig. 2(c). This, together with the large magnetization and the low saturation field, makes it possible for an ideal magnetocaloric material.

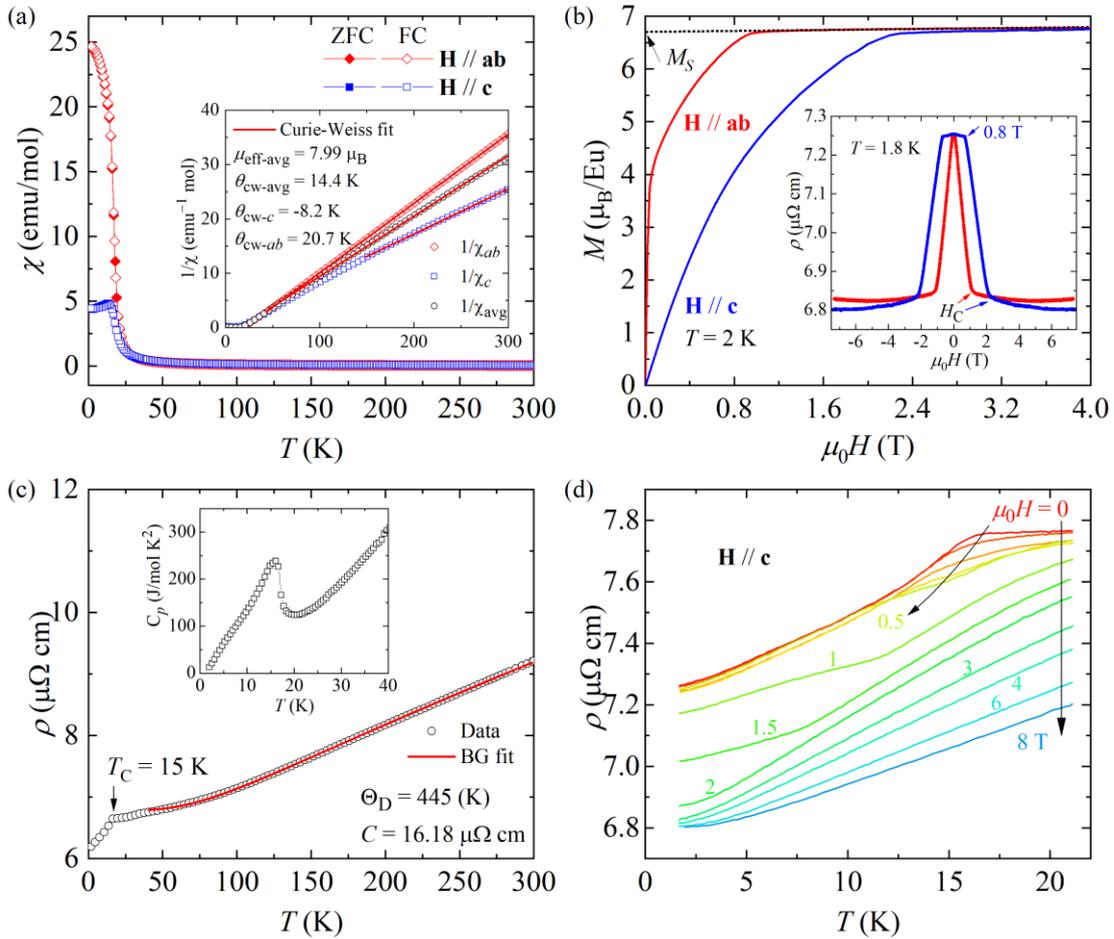

**FIG. 2** (a) Magnetic susceptibility of EuAl$_3$Si after zero-field-cooling (ZFC) and field-cooling (FC) processes under field of 0.1 T. The inset shows inverse susceptibility $\chi^{-1}$ and the Curie-Weiss fittings; the powder-averaged magnetic susceptibility is defined as $\chi_{avg} = (2\chi_{ab}+\chi_c)/3$. (b) Magnetization curves measured at 2 K. The inset shows field dependence of the resistivity at 1.8 K. (c) The temperature dependence of resistivity of EuAl$_3$Si single crystal; the red line represents the fitting using the Bloch-Grüneisen model. The inset displays specific heat as a function of $T$, which shows a λ-shaped peak near $T_C$. (d) $\rho(T)$ profiles measured under various fields **H** // **c**.

To unveil more details about the magnetic order, we performed electrical resistivity ($\rho$) measurements under magnetic field. In general, the $\rho(T)$ presents a typical metallic behavior, and the CDW-induced anomaly seen in EuAl$_4$ is absent here. A remarkable kink is observed at $T_C$ = 15 K due to the reduction of magnetic scatterings, as shown in Fig. 2(c). A residual resistivity $\rho_0$ of 6.19 µΩ cm and a modest residual resistivity ratio (RRR) $\rho$(300 K)/$\rho_0$ of 1.5 can be obtained. The $\rho(T)$ curve can be well fitted on the data from 48 to 300 K using the Bloch-Grüneisen model:[30]

$$\rho(T) = \rho_0 + C\left(\frac{T}{\Theta_D}\right)^5 \int_0^{\frac{\Theta_D}{T}} \frac{x^5}{(e^x-1)(1-e^{-x})} dx, \qquad (1)$$

where $\Theta_D$ is the Debye temperature. The refined parameters are $\Theta_D$ = 445 K and $C$ = 16.18 µΩ cm. Fig. 2(d) shows the temperature dependence of longitudinal resistivity with the **H**//**c** axis for field from 0 to 8 T. As the magnetic field increases, the kink at $T_C$ shifts to lower temperature and fades away, reminiscent of the field effect on AFM-correlated systems. This also suggests that a small AFM component may survive along $c$ axis upon Si doping. However, such AFM correlation seems fragile, and is apt to be polarized by external field. This can be seen from the field dependent resistivity profiles shown in the inset to Fig. 2(b). Generically, negative magnetoresistances are observed at 1.8 K for both field orientations, which again is consistent with FM-dominated ordering. The inflection points in $\rho$(H) signify the field-induced polarization of Eu moments, and the values are 0.8 T and 2.0 T for **H**//**ab** plane and **H**//**c** axis, respectively; both are in good agreement with magnetization isotherms. Notably, an additional inflection point was also recognized for **H**//**c** near 0.8 T, above which the reduction of $\rho_{xx}$(H) greatly speeds up. Combined with the $\rho$(T) curves displayed in Fig. 2(e), we infer that this inflection is probably a consequence of field-induced reorientation from the small AFM component. Altogether, our magnetic and transport measurements manifest that by 25% Si-for-Al doping, the AFM ground state in EuAl$_4$ has been changed into a FM-dominated state, and this encourages us to further look into the MCE in EuAl$_3$Si.

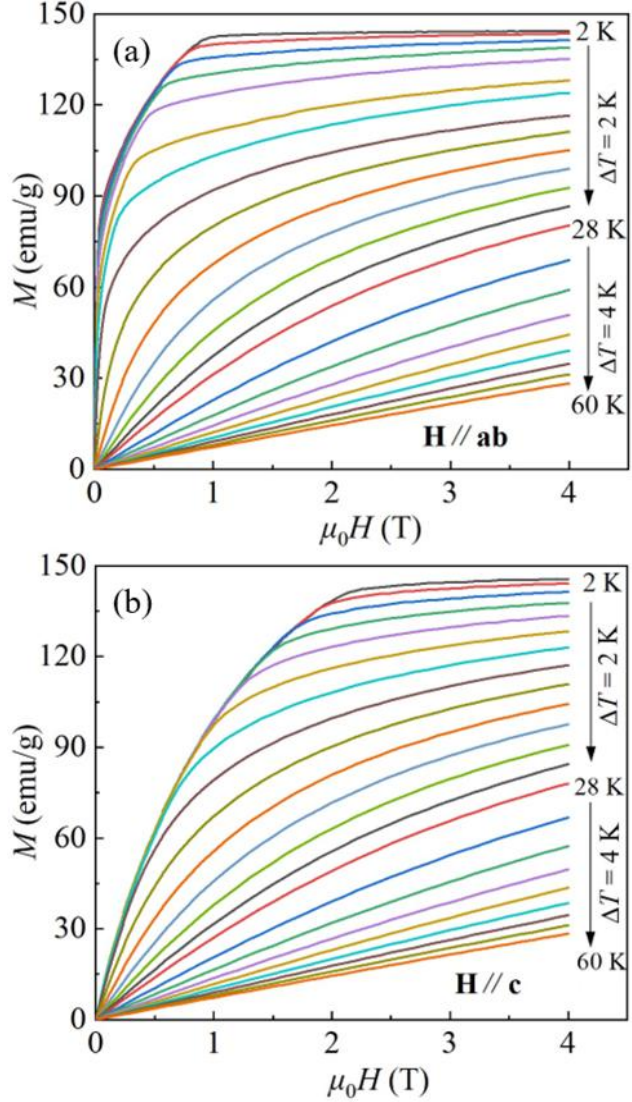

**FIG. 3** Isothermal magnetization curves for field along **ab** plane (a) and **c** axis (b), respectively.

The MCE parameters can be derived by the isothermal magnetization curves, as illustrated in Fig. 3(a) and (b), over the temperature range 2 to 60 K. With temperature rising, the magnetization decreases gradually and $H_C$ moves to a lower value below $T_C$.

The magnetic entropy change [$\Delta S_M (T, H)$] of EuAl$_3$Si was calculated based on the isothermal magnetization around $T_C$ by the Maxwell's relation:[9]

$$\Delta S_M(T,H) = S_M(T,H) - S_M(T,0) = \int_0^H \left[\frac{\partial M(H,T)}{\partial T}\right]_H dH. \qquad (2)$$

The -$\Delta S_M$ as a function of temperature for various magnetic field variations $\Delta H$ are illustrated in Fig. 4(a) and (b). Each curve shows a peak near 15 K, revealing the giant MCE originates from the sharp change of magnetization near FM transition. The peak values of -$\Delta S_M(T)$ reach 13.4 J/kg K and 10.4 J/kg K for the relatively small magnetic field change $\mu_0\Delta H = 2$ T with field along $ab$ plane and $c$ axis, respectively. The greater magnetic entropy is obtained for **H**//**ab**, attributable to the increased availability of free magnetic moments at this orientation. It is noteworthy that the -$\Delta S_M(T)$ curves display a shoulder near 10 K at lower fields along the $ab$-plane. This shoulder grows as field strengthens and finally, the overall peak appears table-like. The observed phenomenon is likely attributable to the Schottky anomaly of the 4$f$-spins in the presence of magnetic exchange field, which is manifested in the specific heat (inset of Figure 2(c)) by a hump around 10 K. Such effect was also observed in EuAl$_4$ and EuGa$_4$ alloys[22,32]. A similar plateau of -$\Delta S_M(T)$ curves was also observed in PrGa, which is closely related to the unconventional magnetic structure, two neighboring magnetic transitions, and the magnetization processes.[33] This plateau-type -$\Delta S_M(T)$ curve has important implications for magnetic refrigeration applications, allowing for the use of a single material to achieve refrigeration over a wider temperature region.

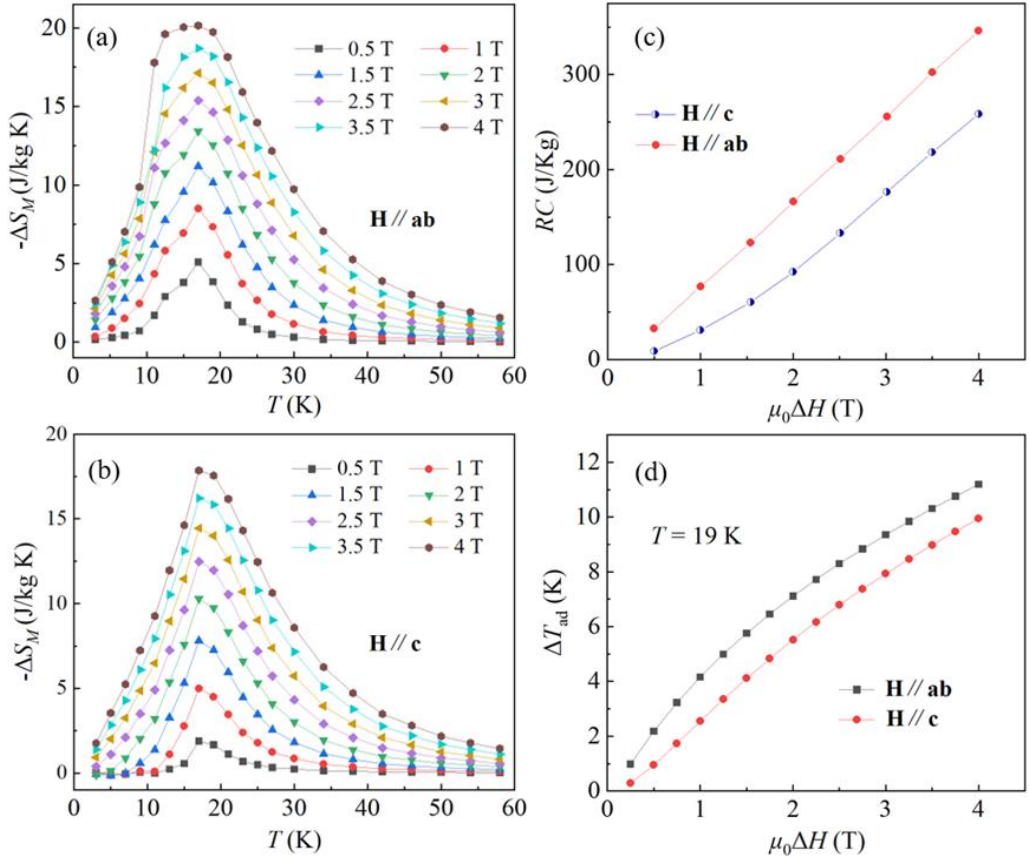

**FIG. 4** The temperature dependence of -$\Delta S_M$ for field along **ab** plane (a) and **c** axis (b) upon different field changes. The $\mu_0\Delta H$ dependence of $RC$ (c) and $\Delta T_{ad}$ at 19 K (d).

Refrigeration capacity ($RC$) is another crucial quantity for assessing the MCE. Following the prevalent definition, $RC$ is determined using the following equation:

$$RC = \int_{T_1}^{T_2} |\Delta S_M|\, dT, \qquad (3)$$

where $T_1$ and $T_2$ are the boundary values of the full width at half maximum in -$\Delta S_M(T)$ curve, respectively. Fig. 4(c) shows the field dependent $RC$ for both directions. Apparently, the $RC$ values increase monotonically with $\Delta H$. The obtained values are 166 J/kg and 92 J/kg at $\mu_0\Delta H$ = 2 T with the field along *ab* plane and *c* axis, correspondingly.

**Table I** The magnetic ordering temperature ($T_M$), $-\Delta S_M$, $RC$, and $\Delta T_{ad}$ under $\mu_0\Delta H$ of 0-2 T and 0-5 T for EuAl$_3$Si, and other typical magnetic refrigeration materials around

the liquid hydrogen temperature. The symbols "-" represent unknown values in the literatures.

| Material | $T_M$ (K) | $-\Delta S_M$ (J/kg K) 2 T | $-\Delta S_M$ (J/kg K) 5 T | RC (J/kg) 2 T | RC (J/kg) 5 T | $\Delta T_{ad}$ (K) 2 T | $\Delta T_{ad}$ (K) 5 T | Reference |
|---|---|---|---|---|---|---|---|---|
| EuAl$_3$Si | 15 | 13.4 | - | 166 | - | 7.2 | - | this work |
| EuAl$_4$ | 15.8 | 8.4 | 21.1 | 67 | 282 | 4.4 | - | 17 |
| EuTiO$_3$ | 5.6 | 22 | 42.4 | 140 | 353 | 10 | 16.6 | 34 |
| EuHo$_2$O$_4$ | 9 | 9 | - | 60 | - | 3 | - | 35 |
| EuO | 69 | 8.4 | 17.5 | - | - | 3.2 | 6.8 | 36 |
| Eu$_3$O$_4$ | 6.5 | - | 12.7 | - | - | - | 7 | 37 |
| GdCoC$_2$ | 15 | 16 | 28.4 | 160 | 369 | - | - | 38 |
| DyNi$_2$ | 20 | 3.1 | 6.2 | - | - | 4.5 | 8.5 | 39 |
| HoB$_2$ | 15 | 24 | 40.1 | - | - | 6 | 12 | 40 |
| HoNi$_2$ | 15 | 14.5 | 25.9 | - | 402 | - | - | 41 |
| ErAl$_2$ | 14 | 21 | 36 | - | - | 5.8 | 11.5 | 42 |
| Er$_3$Ni$_2$ | 17 | 11 | 19.5 | - | - | 3.8 | 5.9 | 43 |
| ErFeSi | 22 | 14.2 | 23.1 | 130 | 365 | 3.4 | 5.7 | 44 |
| TmZn | 8 | 19.6 | 26.9 | - | - | 3.3 | 8.6 | 8 |
| TmGa | 11.5 | 20.6 | 35 | 149 | 364 | 5 | - | 6 |
| TmCuAl | 2.8 | 17.2 | 24.3 | 129 | 371.7 | 4.6 | 9.5 | 12 |
| GdCoC | 14.8 | 20.0 | 35.2 | 59 | 214 | 7.8 | 14.4 | 45 |
| PrZnSi | 13.5 | - | 13.5 | - | 206.5 | - | - | 46 |
| Tm$_6$Ni$_{2.3}$In$_{0.7}$ | 7.6 | 7 | 15 | - | - | - | - | 47 |

A third important parameter of magnetic refrigerant materials is the adiabatic temperature change $\Delta T_{ad}$, which characterizes the temperature increase when a magnetic field is applied adiabatically and can be obtained by the following equation:

$$\Delta T_{ad}(T,H) = -\Delta S_M(T,H)\frac{T}{C_P}, \tag{4}$$

where $C_p$ is specific heat as shown in the inset to Fig. 2(c). The field dependences of $\Delta T_{ad}$ for EuAl$_3$Si at 19 K is shown in Fig. 4(d). For the field change of 0-2 T, the maximum values of $\Delta T_{ad}$ are estimated to be 7.2 K and 5.5 K, respectively.

For comparison, Table I lists the effective parameters of the MCE under $\mu_0\Delta H$ of 0-2 T and 0-5 T for the EuAl$_3$Si (**H//ab**) and other prototypical magnetic refrigeration materials. It is noticeable that the magnetocaloric parameters of EuAl$_3$Si under the low field change are comparable to those of other well-known magnetocaloric materials around 20 K, suggesting it a promising material for hydrogen liquefaction. In particular, there is a significantly improvement of MCE performance in EuAl$_3$Si when compared with its parent compound EuAl$_4$. The -$\Delta S_\mathrm{M}(T)$, $RC$ and $\Delta T_\mathrm{ad}$ parameters are enhanced by 60%, 148% and 64%, respectively, albeit that their magnetic transition temperatures are comparable. Since Si-for-Al doping does not change severely the molar mass or the magnitude of Eu$^{2+}$ moment, the greatly enhanced MCE in EuAl$_3$Si should be due to the modification of the magnetic ground state, as we proposed originally. It is observed that the lattice parameter $a$ remains largely constant compared to the parent compound EuAl$_4$, whereas the parameter $c$ undergoes a significant reduction (11.165 Å → 10.852 Å) upon doping with the greater electro-negative element Si. The distance decrease of the interlayer Eu atoms effectively modulates the RKKY interaction, with the result that it enhances both the ferromagnetic interaction and the temperature at which the magnetic phase transition occurs. The sign of RKKY interaction (FM or AFM) is mainly determined by the factor $\frac{\cos(2k_F r)}{r^3}$, where $r$ is the distance between magnetic moments that can be modified by doping. The typical supporting evidence for this is the ground state of EuAl$_{4-x}$Si$_x$ remains antiferromagnetic-dominated upon 16.25% Si doping with $c$-axis of 10.912 Å.[17] Apparently, varying amounts of Si doping play a significant role in modulating the magnetic ground state of EuAl$_4$ and can enhance the magnetocaloric effect accordingly.

## IV. SUMMARY

In conclusion, we present a systematic investigation into the magnetic properties, magnetotransport characteristics, and magnetocaloric effect of EuAl$_3$Si. Magnetic and transport measurements indicate that EuAl$_3$Si undergoes a ferromagnetic-paramagnetic transition at $T_C$ = 15 K. A giant reversible MCE and large $RC$ have been observed around $T_C$. With a small operation field 2 T, the parameters -$\Delta S_M$, $RC$ and $\Delta T_{ad}$ reach moderate values 13.4 J/kg K, 166 J/kg, and 7.2 K, respectively, which are greatly promoted with respect to the parent compound EuAl$_4$, and are comparable to if not larger than many other prototypical refrigeration materials around liquid hydrogen temperature. These results indicate that EuAl$_3$Si can be a promising candidate to the liquefaction of hydrogen gas. Our work also suggests that modulating the RKKY interaction by proper carrier doping may be constructive to the enhanced MCE in EuAl$_3$Si; the same strategy may be also applicable in pursing higher performance of magnetic-refrigeration materials.

## ACKNOWLEDGMENTS

This research is funded by the National Key R&D Program of China (Nos. 2022YFA1602602 and 2023YFA1609600), the Fundamental Research Funds for the Central Universities (No. YCJJ20230108), the National Natural Science Foundation of China (No. U23A20580), the open research fund of Songshan Lake Materials Laboratory (No. 2022SLABFN27), the Guangdong Basic and Applied Basic Research Foundation (No. 2022B1515120020), and the Beijing National Laboratory for

## DATA AVAILABILITY

The data supporting the findings of this study are available from the corresponding authors upon reasonable request.